\documentclass[twocolumn,prl,aps]{revtex4}

\usepackage{graphicx}
\usepackage{dcolumn}
\usepackage{bm}

\begin{document}

\title{Direct test of pairing fluctuations in the 
pseudogap phase of underdoped cuprates}

\author{N. Bergeal$^1$, J. Lesueur$^1$, M. Aprili$^1$, G. Faini$^2$, J. P. Contour$^3$ and B. Leridon$^1$}

\affiliation{$^1$Laboratoire de Physique Quantique - ESPCI/UPR5-CNRS, 10 Rue Vauquelin - 75005 Paris, France}
\affiliation{$^2$Laboratoire de Photonique et de Nanostructures LPN-CNRS, Route de Nozay, 91460 Marcoussis, France}
\affiliation{$^3$Unit\'e Mixte de Physique CNRS/THALES, (CNRS-UMR137) Route d\'epartementale 128, 91767 Palaiseau Cedex, France}
\date{\today}

\begin{abstract}
We report on a direct test of pairing fluctuations in the pseudogap regime of underdoped superconducting cuprates using a Josephson junction. In this experiment, the 
coupling
between a rigid 
superconducting pair field and pairing fluctuations produces a strong specific signature in the current-voltage characteristics.
Our results show that fluctuations survive only close to T$_{c}$ (T-T$_{c}<$15K) and therefore cannot be responsible for the opening of the pseudogap at high 
temperature.
\end{abstract}
\pacs{74.50.+r 
74.40.+k 
74.72.-h 
}
\maketitle
The normal state of high-temperature superconductors changes qualitatively as a function of the temperature  and the doping 
level of charge carriers, suggesting different ground states \cite{pepin}. In particular, in the underdoped regime, a loss of spectral weight in the electronic excitation spectrum, the so-called pseudogap (PG), is observed above 
T$_{c}$, and below a characteristic temperature T$^*$ \cite{timusk}. The origin of this phenomenon is hotly debated: is it related to superconductivity or to 
a competing hidden order?  Many believe that the answer to
 this question may hold the key to the understanding of high-T$_c$ superconductivity.\\
 \indent  As first observed in the spin channel by NMR, the PG has also been observed by
  most of the one particle probes of electronic excitations \cite{timusk}. Angle Resolved Photoemission Spectroscopy \cite{ding} and Scanning Tunneling 
  Spectroscopy \cite{renner}
  showed that the charge channel is also affected but, more importantly, displayed a characteristic energy of the PG which merges with the superconducting 
  gap when the temperature is lowered below T$_{c}$. Moreover, ARPES data also showed that the pseudogap has the same anisotropy as the superconducting gap 
  in momentum space \cite{norman}. All these observations reveal a smooth crossover rather than a sharp transition line between the pseudogap regime and the 
  superconducting state, and have led to the superconducting precursor scenario. As opposed to the conventional BCS transition where pairing and condensation 
  occur simultaneously at T$_{c}$, in underdoped cuprates fluctuating pairs may form at T$^*$, with no long range coherence, and condense in the superconducting state
   at T$_{c}$ \cite{emery,randeria}. Difficulties in confirming (or invalidating) this scenario arise from the fact that most of the experimental techniques used to investigate the 
   pseudogap are sensitive to the one-particle excitations only, and therefore cannot provide a test of pairing above T$_{c}$. Due to its ability to probe the 
   properties of the superconducting wave function, the Josephson effect is a natural way to address the fluctuation issue.\\
\indent We have designed an original Josephson-like experiment to directly probe the fluctuating pairs in the normal state by measuring the imaginary part
 of the pair susceptibility. In a second order phase transition, the susceptibility is given by the linear response of the order parameter to a suitable external field.
  In the case of the superconducting phase transition, Ferrel and Scalapino showed that the role of the external field could be played by the rigid pair field of a second 
 superconductor below its own T$_{c}$ \cite{ferrel,scalapino}. In a Josephson junction in which one side of the junction is the fluctuating superconductor of interest above its T$_{c}$, 
 while the other side is a superconductor below its T$_{c}$, the coupling between the pairing fluctuations and the well established pair field gives rise to an excess
  current proportional to the imaginary part of the frequency and wave number-dependent pair susceptibility $\chi(\omega,q)$. For a conventional superconductor 
  above its T$_{c}$ \cite{scalapino}
\begin{equation}
 \chi^{-1}(\omega,q)=N(0)\epsilon(1-i\omega/\Gamma_{0}+\xi^2(T)q^2)\ \ \ ,
\end{equation}
where $\Gamma_{0}=(16k_{B}/h)(T-T_c$) is the relaxation frequency of the fluctuations, $\xi(T)$ the superconducting coherence length, $N(0)$ 
the quasiparticle density of states  at the Fermi level and $\epsilon=(T-T_{c})/T_{c}$. The frequency $\omega$ is related to the $dc$ bias voltage $V$ across the junction
 through the Josephson relation $\omega=4¹eV/h$ and the wave number $q$ is related to a magnetic field applied parallel to the junction. In the absence of 
 magnetic field, the excess current is given by \cite{scalapino}
\begin{equation}
 I_{ex}(V)=A\frac{\omega/\Gamma_{0}}{\epsilon[1+(\omega/\Gamma_{0})^2]}\ \ \ ,
\end{equation}
where $A$ depends on the coupling through the barrier and on the characteristics of the superconductors. An explicit calculation has been done
 by Ferrel \cite{ferrel}.\\
\indent Let us emphasize that this dc measurement is sensitive to the pair fluctuations at any frequency (the voltage sets it) and that its temperature dependence is 
mainly controlled by the distance to T$_c$ through $\epsilon$ and $\Gamma_{0}$.
         \begin{figure}[t]
\includegraphics[width=8.5cm,height=4cm]{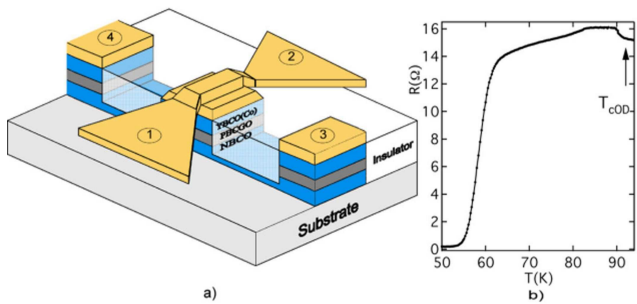}
\caption{\label{fig:epsart1} (Color online) a)  schematic view of a c-axis 
YBCO(Co)(100nm)/ PBCGO(30 or 50nm)/NBCO(200nm) trilayer
 junction. For the sake of clarity, the insulating part in front of the junction is not represented. Each junction is protected by an in-situ gold layer and is connected with two
  electrodes on his top (1 and 2) and two electrodes to the bottom (3 and 4). b) R(T) of a  5$\times$5 $\mu$m$^2$ junction made with a 30 nm thick barrier. At T=90K the optimally doped counter-electrode 
becomes superconducting: this produces a weak current redistribution in the junction (see arrow). Below T$_{cUD}\approx$61K, Josephson coupling occurs and 
the resistance drops.  The finite resistance is due to the gold layer resistance
 (150$m\Omega$) in serial with the junction.}
\end{figure}
 In the 1970Õs, Anderson and Goldman observed gaussian fluctuations just above
 the transition temperature of conventional superconductors \cite{anderson}.They measured an excess $dc$ current through tin/tin oxide/lead junctions
  (for T$_c$(Sn)$<$T$<$T$_c$(Pb)), in qualitative agreement with relation (2). Janko et al, proposed a similar experiment, where the superconductivity of an 
  optimally doped (OD) cuprate is used to probe the superconducting fluctuations in the PG regime of an underdoped (UD) cuprate with a lower T$_c$ \cite{janko}. They
   predicted that an excess current in the junction should persist up to T$^*$ if, according to their model, incoherent pairs are responsible for the PG phase.
    Independently of their respective theoretical framework, all the scenarios involving pairing fluctuations formed at T$^*$ should lead to the same 
    conclusion. On the contrary, for a standard BCS-like transition, the contribution of pairing fluctuations should be limited to the vicinity of T$_{cUD}$.\\
\indent Josephson-like structures involving two different materials  have to be made with thin films. Since high-T$_{c}$ 
compounds grow at high temperature where diffusion is fast, underdoping cannot be obtained by changing the oxygen concentration in one layer only. For the 
coupling to be strong enough, interfaces have to be of very high quality, and therefore an epitaxial structure has to be used: the barrier must have the same 
crystallographic structure as the superconductors.  
Only a few materials can fulfil these requirements. We have chosen : 
 - NdBa$_{2}$Cu$_{3}$O$_{7}$ (NdBCO) as optimally doped compound since it grows smoother than the yttrium compound 
-YBa$_{2}$Cu$_{2.8}$Co$_{0.2}$O$_{7}$ (YBCO(Co)) as underdoped material: Co substitutes Cu in the chains, leading to underdoping with minor disorder
 in the CuO2 
planes \cite{carrington}
 -PrBa$_{2}$Cu$_{2.8}$Ga$_{0.2}$O$_{7}$ (PBCGO) as the barrier: PrBa$_{2}$Cu$_{3}$O$_{7}$ is a weak insulator, and doping with Ga increases
  its resistivity. For this experiment we
  used mainly 30 and 50 nm thick barriers.  
c-axis trilayer structures YBCO(Co)/PBCGO/NdBCO have been grown on SrTiO$_{3}$ (100) substrates by pulsed laser deposition and covered 
by an in-situ gold layer. UV photolithography and ion irradiation have been used to design trilayer junctions whose dimensions range from 40$\times$40$\mu$m$^2$
to 5$\times$5 $\mu$m$^2$ within a wafer (Fig 1) \cite{bergeal}. The details of this completely new process developed on purpose will
 be given elsewhere \cite{bergeal}. Its  reproducibility 
provides the basis for the reliability of these experiments. \\ 
   \begin{figure}[b]
\includegraphics[width=7.5cm]{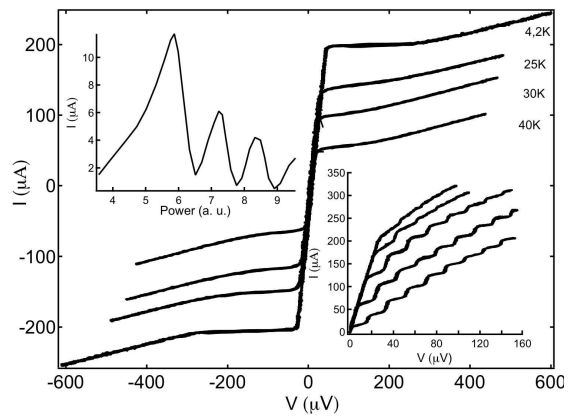}
\caption{\label{fig:epsart1}  Current-voltage characteristics of a  10$\times$10 $\mu$m$^2$ junction made with a 50 nm thick barrier. A T=4.2K, 
I$_c\approx$200$\mu$A and R$_{n}\approx$10$\Omega$. The finite slope of the Josephson current is due to the gold layer resistance
 (150$m\Omega$) in serial with the junction. Critical currents and normal state resistances are found to scale with the surface of the junction. Inset (right): positive part of the 
 I(V) characteristics of the junction under microwave radiations (f=8GHz). Shapiro steps (n=0,1,2,3) appear when the radiation power is increased 
 (from top to bottom). Inset (left): oscillation of the current height of the Shapiro step corresponding to the voltage $V=2ef/h$ ($n$=1).}
\end{figure}
\indent Figure 1 b) displays a typical resistance versus temperature curve of a junction made with a 30 
nm thick barrier. Below T$_{cOD}$=90K, the high resistance of the barrier (15$\Omega$) and the equipotential gold layer on the top of the mesa guarantee that 
the current flows homogeneously along the c-axis in the junction, and that the voltage drop measured in this experiment is dominated by the barrier. 
At T$\approx$61K the UD compound becomes superconducting as expected, and as also observed by measuring its magnetic susceptibility by SQUID 
magnetometry. A Josephson coupling occurs between the two layers and the resistance of the junction drops to zero. 
\indent Before describing the
 temperature regime of main interest  (61K$\rightarrow$90K), we first establish that both $dc$ and $ac$ Josephson effects do occur when both electrodes are in the
  superconducting state.  This is of great importance since the excess current in the fluctuating regime has the same origin as the Josephson one at low 
  temperature. Below the two T$_{c}$, current-voltage characteristics display a typical Josephson RSJ-like behaviour with an I$_c$R$_n$ product of 2 meV at 
  4.2K (Fig 2). The current-voltage characteristics exhibit clear Shapiro steps at fixed voltage  $V_{n}=n2ef/h$ (n=0, $\pm$1, $\pm$2É) when the junction is
   irradiated with microwaves of frequency $f$ (Fig 3 right inset). We have checked that the width of the steps satisfies the linear relation with frequency and that
    the current height of the steps modulates with the microwave power as expected (Fig 3 left inset) \cite{barone}. Such a Josephson effect through PBCO (or PBCGO) 
    barriers has been reported by several groups. This material is known to contain localised states which control the transport; the Josephson effect takes 
    place by direct or resonant tunneling through localized states in the barrier 
    \cite{golubov,bari,devyatov97}. At finite energy, quasiparticle transport occurs by hopping through these states \cite{yoshida,glazman}. In our junction, 
    the background conductance of a 
    30 nm thick barrier for T$>$T$_{cOD}$ has a weak dependence with energy, as expected for one or two localized states in the barrier. The conductance follows 
    the characteristic law $G=G_{0}+\alpha V^{4/3}$ while junctions with a 50 nm thick barrier exhibit the power law $G=G_{0}+\alpha V^{4/3}+\beta V^{5/2}$ expected
     for three localized states \cite{yoshida}. Since the transport includes non-elastic hopping, no clear 
     spectroscopic signatures are expected compared to tunnel junction with conventional superconductors. It must be stressed that tunneling is not a 
     requirement for this Josephson-like experiment. Doping PBCO with Ga reduces the number of localized states, and allows to use barriers with a very few of them,
      but thick enough to avoid microshorts due to extrinsic inhomogeneities.\\ 
\indent We now focus on the intermediate regime of temperature (T$_{cUD}$$<$T$<$T$_{cOD}$), the one of main interest here. In order to increase the sensitivity
 of the experiment, we measure the dynamic conductance $G=dI/dV$ of the junction as a function of the bias voltage. Figure 3 a) displays
  typical results. At high energy ($>$10meV) the background exhibits a weak dependence with energy and temperature indicating that the quasiparticle
  transport is dominated mainly by one or two localized states. In addition to the quasiparticle background, an excess conductance peak emerges from the 
  Josephson current at zero energy when the temperature crosses T$_{cUD}$, and reduces rapidly when the temperature is increased further. It disappears 14K above 
  T$_{cUD}$, well below the characteristic temperature expected for the PG in our compound (T$^*\approx$250K estimated from resistivity measurement),  as opposed to Janko's
   prediction. The peak presents all the characteristics
   expected from standard gaussian fluctuations above T$_{cUD}$, as calculated and observed in conventional superconducting transitions \cite{anderson}.\\
   \indent Figure 3 b) displays 
    the result of
     $G_{ex}$  calculation as a function of temperature, in qualitative agreement with the experimental data, both in energy and amplitude, provided the phase
      fluctuations introduced by voltage noise in actual junctions is properly taken into account. This is simply due 
    to Johnson noise in this rather high temperature experiment. Following reference \cite{kadin}, $\Gamma_{0}$ has to be replaced by $\Gamma=\Gamma_{0}+\Gamma_{1}$
     where $\Gamma_{1}=4e^2Rk_{B}T/\hbar^2$ and $R$ is the resistance of the 
    junction. As expected, thermal noise cuts  the low energy part of the fluctuation spectrum.\\
The excess conductance peak is  strongly 
     suppressed by microwave radiation ; this can be used to get suitable background to extract the excess conductance due to fluctuations. The result is shown in the  
     inset figure 3 b) : the overall shape of the curve is in good agreement with the calculation. The energy
      scale $h\Gamma /4eV$(voltage corresponding to $G_{ex}$=0) is close to the expected value from the gaussian model, albeit a little larger. This
       is an important issue, since $\Gamma$ is related to the characteristic relaxation frequency 
     of the fluctuations. The small discrepancy may originate from the details of the transport (localized states, d-wave symmetryÉ), and the 
     actual choice of $T_{cUD}$, since 
     there is a finite transition width. As microwaves may also affect the quasiparticles transport, this method is not accurate enough to make a full quantitative comparison
     of the $G_{ex}$ data with the model, and to determine $\Gamma$ for each temperature. \\
                       \begin{figure}[t]
\includegraphics[width=8cm]{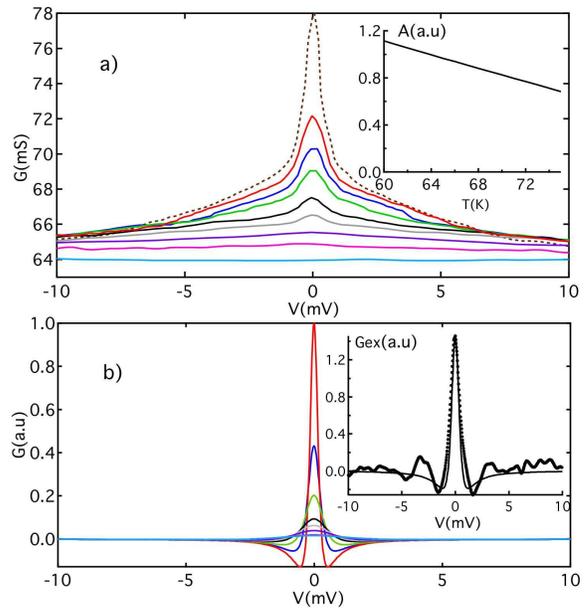}
\caption{\label{fig:epsart1} (Color online) a) Dynamic conductance as a function of energy of a 5$\times$5 $\mu$m$^2$junction made with a 30 nm thick barrier. The curves
 correspond to the following temperatures (from top to bottom) : T-T$_{cUD}$=$\delta$ (dotted line), 2.5K, 4K, 6K, 9K , 11K , 14K , 20K , 22K. The excess
  conductance peak 
 emerges from the junction current at T$_{cUD}$ (T=T$_{cUD}$-$\delta$) and disappears at T-T$_{cUD}\approx$14K. Inset  :
  computed temperature dependence of the coupling factor A according to Ferrel \cite{ferrel}.
b) Computed
  excess conductance curves using the Kadin and Goldman model for noise \cite{kadin},  for the same temperatures
   (T- T$_{cUD}$ from 2.5K to 22K) ($R= 15\Omega$, C=1.4 10$^{-14}$ F).  Inset : Excess conductance obtained by subtracting the conductance under microwaves  (dots)
   compared with the calculation at T-T$_{cUD}$=6K (solid line).} 
\end{figure}
     \indent The data do not follow Janko's predictions. The extra-contribution due 
to fluctuating pairs is expected to move towards high energy when the 
temperature is increased (typically 10meV at $T/T_{cUD}$=1.1) : this 
is not observed here.  The broad feature, which extends up to 10 meV 
      is seen in all the samples but can not be attributed to fluctuations since it is already present at low temperature (where no fluctuation are present) and evolves continuously through $T_{cUD}$
       up to $T_{cOD}$,
       where it disappears. Following Devyatov et al \cite{devyatov98}, we therefore attribute this feature to Andreev Reflection in the presence of localized states. G(V) is highly symmetric and doesn't
       show any evidence of the particle-hole asymmetry suggested by Janko.\\ 
\indent The temperature dependence of the pairing peak is a key point. We have already ruled out the barrier conductance itself as a possible explanation for the 
temperature dependence of the observed signal. We can exclude any contribution of the UD layer itself since the c-axis conductance of underdoped YBCO 
increases with temperature in this temperature range \cite{takenaka}, and since no specific energy dependence is expected. The temperature dependence of the OD layer 
properties has to be addressed. It must be emphasized that in this linear response experiment, the current is directly proportional to $Im\chi$, which is independent
 of the strength of the external field. The full calculation of the linear coefficient $A$ shows that the excess current is weakly sensitive to the temperature dependence
  of the fully established superconducting pair field in this temperature range (see Fig 3 a) inset). This contribution has been properly taken into account in the
   present calculation. Therefore, the decrease of the excess conductance peak cannot be trivially attributed to the reduction of the superconducting pair field in the
    optimally doped layer. The excess current is observed in the temperature range where gaussian fluctuations are expected in cuprates, i.e. roughly 15K above 
    T$_{cUD} $given their short coherence length and the weak anisotropy of YBCO. As an example, the Lawrence-Doniach calculation of the paraconductivity above
     T$_c$ leads to less than 5$\%$ of excess conductivity in this range of temperature.\\
\indent The strength of our
       experiment is that it relies only on the presence of pairing fluctuations and the Ginsburg-Landau theory. An attempt to indirectly detect pairing above T$_{cUD}$
        by high frequency measurements has been made previously in a restricted range of frequency, and within a precise theoretical framework \cite{corson}. Fluctuations
         have been observed up to 95K only, far below T$^*$, in rather good agreement with our result. Our broadband measurement reveals the presence of gaussian
          fluctuations with no further theoretical assumption.\\
\indent In summary, by probing the pairs directly,  we have found that in the normal phase of an underdoped cuprate, pairing fluctuations survive only in a 
reduced regime of temperature above T$_{cUD}$, which is consistent with a standard model of gaussian fluctuations. This result is in contradiction with the precursor
 superconducting scenario as an explanation of the pseudogap. Consequently, focusing on alternative scenarios seems
  to be a reasonable approach to investigate the underdoped regime of superconducting cuprates.\\
\indent The authors acknowledge M. Grilli, S. Caprara, C. Castellani, C. Di Castro for stimulating discussions. They also thank P. Monod, O. Kaitasov and C. Dupuis.

\end{document}